\documentstyle[12pt,epsfig]{article}
\begin{document}

\begin{titlepage}
\rightline{March 2012}
\vskip 2cm
\centerline{\Large \bf
Mirror dark matter interpretations   
}
\vskip 0.5cm
\centerline{\Large \bf 
of the DAMA, CoGeNT and CRESST-II data}

\vskip 2.2cm
\centerline{R. Foot\footnote{
E-mail address: rfoot@unimelb.edu.au}}

\vskip 0.7cm
\centerline{\it ARC Centre of Excellence for Particle Physics at the
Terascale,}
\centerline{\it School of Physics, University of Melbourne,}
\centerline{\it Victoria 3010 Australia}
\vskip 2cm
\noindent
The CRESST-II collaboration have announced evidence for the direct
detection of dark matter 
in $730$ kg-days exposure of a CaWO$_4$ target.
We examine these new results, along with DAMA and CoGeNT data, in the
context of 
the mirror dark matter framework. We show that all three experiments can
be simultaneously
explained via kinetic mixing induced elastic scattering of a mirror
metal component off target nuclei.
This metal component can be as heavy as Fe$'$ if the
galactic rotational velocity is relatively low: $v_{rot}
\stackrel{<}{\sim} 220$ km/s. 
This explanation is consistent 
with the constraints from the other experiments, such as CDMS/Ge,
CDMS/Si and 
XENON100 when modest $\sim 20-30\%$ uncertainties in energy scale are
considered.

\end{titlepage}


\section{Introduction}

Over the last decade or so, progress has been made in efforts to
directly detect dark matter.
In particular, the DAMA/NaI \cite{dama1} and DAMA/Libra \cite{dama2}
experiments 
have obtained very exciting results in their dark matter search.
Recall that these experiments have observed a modulation 
in the `single hit' event rate with a period and phase consistent with
expectations 
from dark matter interactions\cite{dm}. Background rates are expected to
be time independent
with the possible exception of muon induced backgrounds. However it has
been known
for a long time that muons cannot mimic the dark matter annual
modulation
signature\cite{damamuon}. The DAMA experiments thus provide very
convincing evidence
that dark matter interactions have been detected.
  
Another interesting development is the results obtained in the CoGeNT
experiment\cite{cogent}.
That experiment 
features a Germanium target and very low energy threshold. The CoGeNT
team
observed a rising event rate at low energies. These events could not be
explained
by known backgrounds and can be interpreted as evidence supporting the
DAMA signal.
Very recently,
the CRESST-II collaboration have announced results
for their dark matter search with $730$ kg-days of net exposure in a 
CaWO$_4$ target\cite{cresst-II}.  The CRESST-II data also cannot be
explained by 
known backgrounds and are 
compatible with a dark matter signal also rising at low energies.

Attempts to explain the positive signals of DAMA, CoGeNT and CRESST-II
in terms
of standard WIMP dark matter have proven challenging\cite{std}. 
However, it has been
known for some time\cite{foot69,footold,foot2008}, that mirror dark
matter  
has a number of novel and distinctive features which make it a  suitable
candidate to explain
the data. In particular mirror dark matter 
features a mass dependent velocity dispersion $v_0 \propto {1 \over
\sqrt{m_{A'}}}$, 
$E_R$ dependent Rutherford scattering ${d\sigma \over dE_R}
\propto {1 \over E_R^2}$, and particles
necessarily in the relevant low mass range. With these features the DAMA
and CoGeNT data can be
simultaneously explained\cite{foot2,foot1}.  
The main purpose of this paper is to examine the implications of the new
CRESST-II results for mirror dark matter.
In the process we will update the CoGeNT analysis taking into account
their recently reported
surface event correction factor\cite{collartaupxxx}.  It turns out that
these developments point to new parameter regions
within the mirror dark matter framework. Heavy mirror elements, $\sim$
Fe$'$, can explain all three
experiments if the galactic rotational velocity is low $v_{rot}
\stackrel{<}{\sim} 220$ km/s.
Other regions of parameter space are possible. 
We explore a few examples in some detail and examine the question of
their compatibility with constraints
from sensitive but high threshold experiments such as XENON100 and
CDMS/Ge.

\section{Mirror dark matter and its direct detection}

Mirror dark matter posits that the inferred dark matter in the Universe
arises from
a hidden sector which is an exact copy of the standard model
sector\cite{flv} 
(for a review
and more complete list of references see ref.\cite{review})\footnote{
Note that successful big bang nucleosynthesis and successful
large scale structure requires effectively asymmetric initial
conditions in the early Universe, $T' \ll T$ and $n_{b'}/n_b \approx 5$. 
See ref.\cite{some} for further discussions.}.
That is, 
a spectrum of dark matter particles of known masses are predicted: e$'$,
H$'$, He$'$, O$'$, Fe$'$,... (with
$m_{e'} = m_e,\ m_{H'} = m_H,$ etc). 
Kinetic mixing of the $U(1)_Y$ and its mirror counterpart 
allows ordinary and mirror particles to interact with each
other\cite{he}.
This $U(1)_Y$ kinetic mixing induces
photon-mirror photon kinetic mixing:
\begin{eqnarray}
{\cal L}_{mix} = \frac{\epsilon}{2} F^{\mu \nu} F'_{\mu \nu}
\label{kine}
\end{eqnarray}
where $F_{\mu \nu}$ is field strength tensor for the photon
and $F'_{\mu \nu}$ is the field strength tensor for the mirror photon.
This interaction enables charged mirror sector particles of charge $e$
to couple to
ordinary photons with electric charge $\epsilon e$ \cite{holdom}. 
The cross-section of such a particle, say a mirror nucleus, $A'$, with
atomic
number $Z'$ and velocity $v$
to elastically scatter off an ordinary nucleus (presumed at rest with
mass 
and atomic numbers $A,\ Z$) is given by\cite{foot69}:\footnote{Unless
otherwise indicated, we 
employ natural units where $\hbar = c = 1$.}
\begin{eqnarray}
{d\sigma \over dE_R} = {\lambda \over E_R^2 v^2}
\label{cs}
\end{eqnarray}
where 
\begin{eqnarray}
\lambda \equiv {2\pi \epsilon^2 Z'^2 Z^2 \alpha^2 \over m_A} F^2_A
(qr_A) F^2_{A'} (qr_{A'})   \
\end{eqnarray}
and $F_A (qr_A)$ [$F_{A'} (qr_{A'})$] is the form factor which
takes into account the finite size of the nucleus [mirror nucleus].
A simple analytic expression for
the form factor, which we adopt in our numerical work, is the one
proposed by Helm\cite{helm,smith}.

What about the distribution of mirror particles in the galactic halo?
These are assumed to be, predominately,\footnote{
There may also be a remnant disk of dark mirror stars\cite{macho1}, but
constraints from
gravitational microlensing (MACHOs searches) limit this component to be
subdominant\cite{macho}.}
roughly spherically distributed in a multi-component plasma containing 
e$'$, H$'$, He$'$, O$'$, Fe$'$,...\cite{sph}.  The interaction length is
typically much less than a 
parsec\cite{footz} and the dark matter particles form a
pressure-supported halo, ideally described
by a common temperature, $T$
\footnote{
There are limits on self interactions of dark matter from 
observations of the Bullet cluster\cite{bullet}.
This sets stringent limits on self interactions provided that the bulk
of the dark matter particles are distributed throughout the
cluster and not bound to individual galaxies. 
However mirror dark matter is dissipative
and in clusters (or at least in some of them)
the bulk of the dark matter particles might be confined in galactic
halos [c.f. ref.\cite{zurab}]. 
Under this assumption mirror
dark matter is consistent with Bullet cluster observations.
}.  This temperature can be estimated from
the condition of hydrostatic equilibrium which implies:
\begin{eqnarray}
T = {1 \over 2} \bar m v_{rot}^2 
\end{eqnarray}
where $v_{rot}$ is the galactic rotational velocity and
$\bar m = \sum n_{A'} m_{A'}/\sum n_{A'}$ is the mean mass of the
particles in the halo.
The halo distribution 
of mirror species is then
$f_{A'} = exp(-E/T) = exp(-{1 \over 2} m_{A'} v^2/T) = exp(-v^2/v_0^2)$.
Evidently 
\begin{eqnarray}
v_0[A'] &=& \sqrt{{2T \over m_{A'}}} \nonumber \\
        &=& v_{rot} \sqrt{{\bar m \over m_{A'}}}\ . 
\label{v0}
\end{eqnarray} 
The parameter $\bar m$ can be estimated
within the mirror dark matter model. 
For a H$'$, He$'$ mass dominated halo, taking into account that the
plasma is expected to be fully ionized,
we have
\begin{eqnarray}
{ \bar m \over m_p} \simeq {1 \over 2 - {5 \over 4} \xi_{He'}}\ 
\label{barm2}
\end{eqnarray}
where $\xi_{He'}$ is the mass fraction of He$'$.
The primordial He$'$ abundance can be computed as a function of the
kinetic mixing 
parameter, $\epsilon$,
and for $\epsilon \sim 10^{-9}$ the calculations suggest that the
primordial $He'$ mass fraction, $Y'_p$, 
is around $0.9$\cite{paolo2}.
If $\xi_{He'} \approx Y'_p$ then
Eq.(\ref{barm2}) suggests that $\bar m \approx 1.1 $ GeV. 
Even if mirror stellar evolution depleted H$'$ further
(or somehow depleted both H$'$ and He$'$),
this would have only a relatively minor affect on the estimation of
$\bar m$.

A key feature of this model is that the heavy components with $m_{A'}
\gg \bar m$ 
have $v_0[A'] \ll v_{rot}$. This can drastically reduce the tail of the
distribution [c.f. standard 
WIMPs which have $v_0 = v_{rot}$].
This feature can help explain why higher threshold experiments such as
CDMS/Ge and XENON100 currently do not
see a signal while the lower threshold DAMA and CoGeNT experiments do.

The differential scattering rate for $A'$ elastic scattering on a target
nuclei, $A$, is given 
by\footnote{The upper limit of integration in Eq.(\ref{55}) 
is taken as infinity since we are dealing with
dark matter particles with significant self interactions. The self
interactions will prevent
particles in the high velocity tail of the Maxwellian distribution from
escaping the galaxy. }:
\begin{eqnarray}
{dR \over dE_R} = N_T n_{A'} 
\int^{\infty}_{|{\textbf{v}}| > v_{min}}
{d\sigma \over dE_R}
{f_{A'}({\textbf{v}},{\textbf{v}}_E) \over k} |{\textbf{v}}| d^3v 
\label{55}
\end{eqnarray}
where the integration limit, $v_{min}$, is given by the kinematic
relation:
\begin{eqnarray}
v_{min} &=& \sqrt{ {(m_{A} + m_{A'})^2 E_R \over 2 m_{A} m^2_{A'} }}\ .
\label{v}
\end{eqnarray}
In Eq.(\ref{55}), $N_T$ is the number of target nuclei per kg of
detector and   
$n_{A'}$ is the number density of halo dark matter particles, $A'$, at
the Earth's
location. This number density can be expressed in terms of the halo
mass fraction of species $A'$ , $\xi_{A'}$, and total mass density, 
$\rho_{dm}$ via $n_{A'}  = \rho_{dm} \xi_{A'}/m_{A'}$ 
(we fix $\rho_{dm} = 0.3 \  {\rm GeV/cm}^3$).
Also in Eq.(\ref{55}) ${\bf{v}}$ is the velocity of the halo particles
relative to the
Earth and ${\bf{v}}_E$ is the
velocity of the Earth relative to the galactic halo\footnote{In all
numerical work
we include an estimate of the Sun's peculiar velocity so that  
$\langle v_E \rangle = v_{rot} + 12$ km/s.}.
The halo distribution function, in the reference frame of the Earth, is
then given by
the Maxwellian distribution:
\begin{eqnarray} 
{f_{A' } ({\bf{v}},{\bf{v}}_E) \over k} = (\pi v_0^2[A'])^{-3/2}
exp\left({-({\bf{v}}
+ {\bf{v}}_E)^2 \over v_0^2[A']}\right)\ . 
\label{pw}
\end{eqnarray}
The integral, Eq.(\ref{55}), can easily be evaluated in terms
of error functions\cite{foot2008,smith} and numerically solved.

To compare with the measured event rate, we must include detector
resolution effects 
and overall detection efficiency, $\epsilon_f (E_R^m)$, when the latter
is not already included in the
experimental results:
\begin{eqnarray}
{dR \over dE_R^m} = \epsilon_f (E_R^m) {1 \over \sqrt{2\pi}\sigma_{res}
} 
\int {dR \over dE_R} e^{-(E_R - E_R^m)^2/2\sigma^2_{res}} dE_R 
\label{556}
\end{eqnarray}
where $E_R^m$ is the measured energy and $\sigma_{res}$ describes the
resolution.
The measured energy is typically in keVee units
(ionization/scintillation energy). For nuclear recoils
in the absence of any channeling, keVee = keV/$q$, where $q < 1$ is the 
relevant quenching factor. Channeled events, where scattered target
atoms travel down crystal
axis and planes, have $q \simeq 1$. In light of recent theoretical
studies\cite{newstudy}, we assume
that the channeling fraction is negligible. Channeling, though, is a
complicated
subject and the modeling used in ref.\cite{newstudy} might miss
important physics. 
Therefore it is, of course, still possible that
channeling could play an important role, which could modify the favored
regions of 
parameter space somewhat.

\section{Mirror dark matter interpretation of CRESST-II, DAMA/Libra and
CoGeNT signals}

\vskip 0.5cm
\noindent
{\large \it The CRESST-II experiment}
\vskip 0.5cm

The CRESST-II data\cite{cresst-II} arises from 8 detector modules, with 
recoil energy thresholds (keV) of $10.2, 12.1, 12.3, 12.9, 15.0, 
15.5, 16.2, 19.0$.
The data is plotted in figure 1.
With a view to performing a standard $\chi^2$ analysis of this data,
we bin this data into 5 bins, summarized in table 1. This table
also indicates the expected background rate estimated from all known
sources of 
background\cite{cresst-II}.
\begin{table}
\centering
\begin{tabular}{c c c}
\hline\hline
Bin / keV & Total events & Estimated background  \\
\hline
10.2 -- 13.0 & 9 & 3.2  \\
13 -- 16 & 15 & 6.1 \\
16 -- 19 & 11 & 7.0  \\
19 -- 25 & 12 & 11.5  \\
25 -- 40 & 20 & 20.1 \\
\hline\hline
\end{tabular}
\caption{CRESST-II data: total number of events and estimated
background.}
\end{table}
\vskip 0.9cm

\vskip 0.5cm
\centerline{\epsfig{file=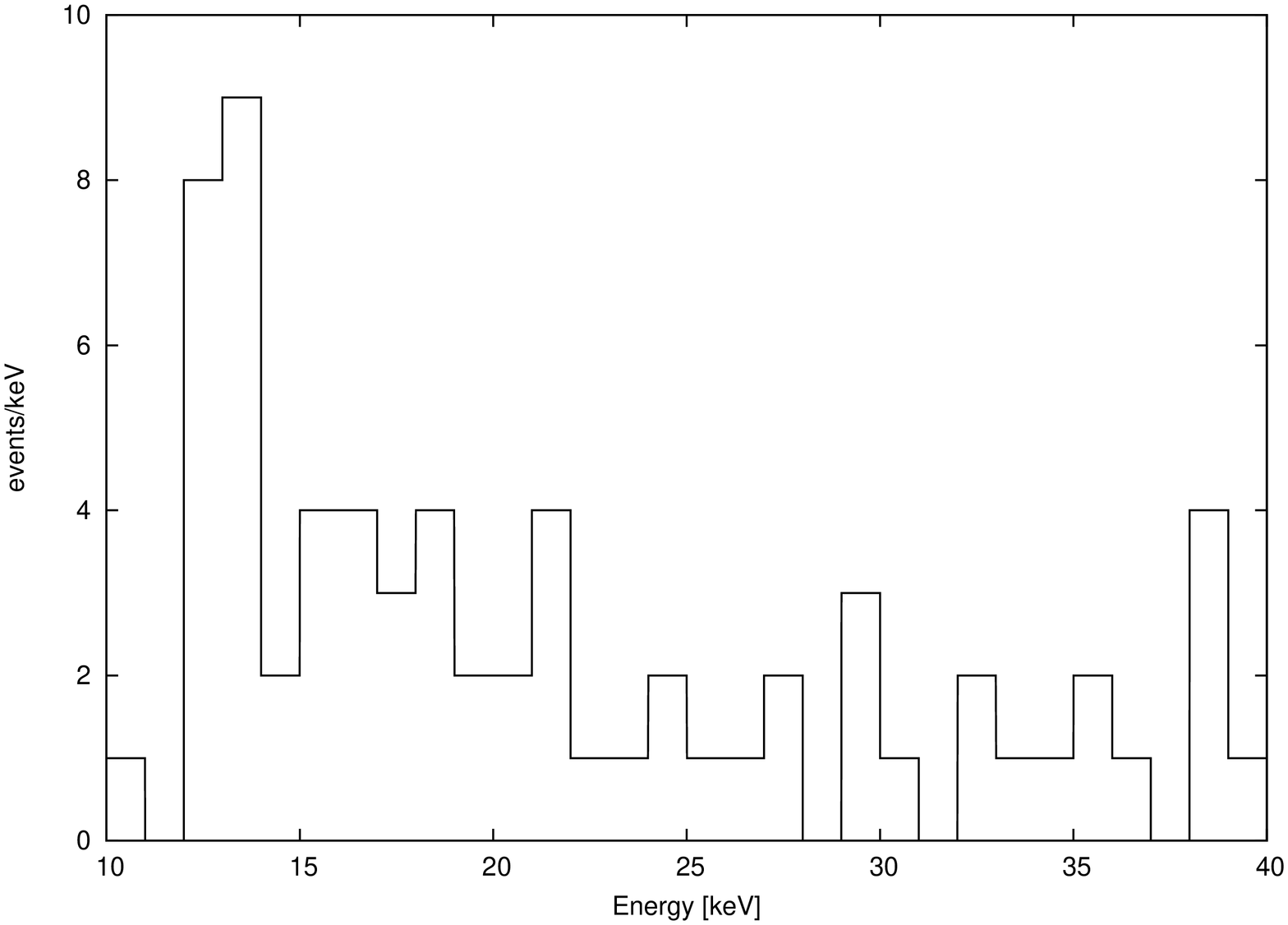,angle=270,width=13.2cm}}
\vskip 0.3cm
\noindent
{\small
Figure 1: CRESST-II data.}
\vskip 0.9cm

The rates, $R_i$, relevant for the CRESST-II experiment are computed 
using Eq.(\ref{55}) and Eq.(\ref{556}). 
As discussed earlier, the halo distributions in Eq.(\ref{55}) depend on
the parameter $\bar m$
which has been estimated to be $\bar m \approx 1.1$ GeV. A reasonable
uncertainty in $\bar m$ is $\pm 20\%$, but such a variation does not
significantly
affect any of our conclusions. We therefore fix $\bar m = 1.1$ GeV in
all of our
numerical work.
We use $\sigma_{res}  = 0.3$ keV\cite{cresst-II}
and assume detection efficiencies for the three target elements:
$\epsilon_f = 0.90$ for O, W and 
$\epsilon_f = 1.0$ for Ca,
which take into account their acceptance region. 
As per previous analysis\cite{foot2,foot1} we make the 
simplifying assumption that the mirror metal components
are dominated by a single element $A'$ in a given experiment. This can
only be an approximation, 
however it can often
be a reasonable one given the relatively narrow energy range probed in
the 
experiments [e.g. the signal region in the experiments are mainly:
2-4 keVee (DAMA), 0.5-1 keVee (CoGeNT),  12-14 keV (CRESST-II)].
With this assumption and assuming a particular value for $v_{rot}$, the
fit to the data depends on two 
parameters
$m_{A'}, \ \epsilon\sqrt{\xi_{A'}}$.
\footnote{In our numerical work we allow $A', Z'$ to have non-integer
values,
with $Z' = A'/2$ (except when we specifically consider $A' = $ Fe$'$,
then 
$Z' = 26$ and $A' \simeq 56$).  
Since the realistic case involves a spectrum of elements, the 
effective mass number can be non-integer.}
We define $\chi^2$ for the CRESST-II data in the usual way:
\begin{eqnarray}
\chi^2 (m_{A'}, \epsilon \sqrt{\xi_{A'}}) = 
\sum_{i=1}^{5}  \left[ {R_i + B_i - data_i \over \delta data_i}\right]^2
\ 
\label{chi2bla}
\end{eqnarray}
where $R_i$ is the theoretically predicted rate and $B_i$ is the
estimated
background in the $i^{th}$ energy bin. We evaluate $\chi^2$ as per
Eq.(\ref{chi2bla}),
with the constraint $m_{A'} \le m_{Fe} \simeq 55.8m_p$.
We have not considered any uncertainty in the CRESST-II energy scale.

\vskip 0.6cm
\noindent
{\large \it  The DAMA annual modulation signal}
\vskip 0.3cm

For DAMA we analyse the annual modulation signal  
using the 12 bins of width $0.5$ keVee in the energy range: $2-8$
keVee\cite{dama2}
taking into account the detector resolution\cite{damares}.
An important issue is the nuclear recoil energy scale which is set by
the values 
of the quenching factors $q_{Na}, \ q_I$.
In the recent analysis\cite{foot2,foot1}
we used $q_{Na} = 0.30 \pm 0.06$, $q_I = 0.09\pm 0.02$, where the
central values
were taken from averages over a large range of energies\cite{dama1}. 
However quenching factors are in general
energy dependent and many measurements of $q_I$ tend to show increasing 
values in the important low energy region. This is also supported by
some theoretical arguments based on the classical Birks forumula\cite{tretyak}. 
[See e.g. ref.\cite{kims} for 
a summary figure of much of the available information].
In view of the current situation, we 
allow for a range of possible values for the quenching factors:
\begin{eqnarray}
q_{Na} = 0.28 \pm 0.08, \ \ 
q_I = 0.12\pm 0.08 
\label{moomba}
\end{eqnarray}
We minimize $\chi^2$ varying  $q_{Na}, \ q_I$ over the above range of
values
(assumed energy independent for simplicity).
Values of $q_I, \ q_{Na}$ outside the above range are certainly possible, 
such as the higher values $q_{Na} \approx 0.6, \ q_I \approx 0.3$
suggested by Tretyak\cite{tretyak}.  While we don't specifically consider
the possibility of such high quenching factors in the numerical
work presented here,
qualitatively we note that higher quenching factors generally move
the DAMA allowed region to lower values of $\epsilon \sqrt{\xi_{A'}}, \ m_{A'}$ 
and also yields interesting parameter space overlapping with that from CoGeNT and CRESST-II.

\vskip 0.3cm
\noindent
{\large \it The CoGeNT experiment}
\vskip 0.3cm

Our analysis procedure for CoGeNT is different to our earlier
analysis\cite{foot2,foot1}.
It has been pointed out that there may be surface events contaminating
CoGeNT's signal region\cite{collartaupxxx}  which have not been
excluded by CoGeNT's rise time cut. Preliminary estimates indicate
that this can be a very large effect - 
the signal
may be reduced by a factor of around $0.3$ in the important low energy 
region ($E_R < 1.0 $ keVee)\cite{collartaupxxx,hooper}.
In view of this we analyse CoGeNT data using 15 bins of 
width $0.1$ keVee in the region $0.5-2.0$ keVee. 
We take the efficiency corrected and surface event corrected CoGeNT data
from
figure 4 of ref.\cite{hooper}.
We allow a constant background, which we fit to the data in this energy
range.
We have taken into account some uncertainties in energy scale by
minimizing the $\chi^2$ for
CoGeNT over the variation in quenching factor, 
$q_{Ge} = 0.21 \pm 0.04$. 
Obviously, the CoGeNT spectrum 
is quite uncertain at the present time and one could easily assign $\sim
40\%$ 
uncertainty in the rate. Such an 
uncertainty would imply a $\sim 20\%$ uncertainty in inferred values of
$\epsilon\sqrt{\xi_{A'}}$
from CoGeNT data.  This is in addition to the uncertainties we have
considered.
Thus, the reader should
be aware that the favored region we derive for CoGeNT might move
considerably as
future data is accumulated and the backgrounds are better understood.

\vskip 0.3cm
\noindent
{\large \it The analysis}
\vskip 0.3cm
\begin{table}
\centering
\begin{tabular}{c c c}
\hline\hline
Experiment &  $\chi^2 (min)/d.o.f. $ & Best fit parameters  \\
\hline
DAMA (annual mod.) &   $5.7/10$ &  ${m_{A'} \over m_p} = 55.8$ , 
\ $\epsilon \sqrt{\xi_{A'}}/10^{-10} = 2.5 $  \\
CoGeNT (spectrum) &    $15.7/12$  & ${m_{A'} \over m_p} = 42.0$ , 
\ $\epsilon \sqrt{\xi_{A'}}/10^{-10} = 1.7 $  \\
CRESST (spectrum) &   $2.6/3$ &  ${m_{A'} \over m_p} = 55.8 $ , 
\ $\epsilon \sqrt{\xi_{A'}}/10^{-10} = 2.7 $  \\
\hline\hline
\end{tabular}
\caption{Summary of $\chi^2 (min)$ for the relevant data sets from the 
DAMA, CoGeNT and CRESST-II experiments. [Here $v_{rot} = 200$ km/s].}
\end{table}

Table 2 gives the $\chi^2_{min}$ values obtained from the fit to
the DAMA annual modulation and CoGeNT, CRESST-II spectrum data for 
a fixed value of $v_{rot} = 200$ km/s, which we have found as an example
where all three experiments have overlapping favored regions of
parameter space.
The value $v_{rot} = 200$ km/s is around $20\%$ less than some recent
estimates of this quantity\cite{rot}. However allowing for a small bulk
halo rotation, e.g. in a co-rotating
halo, a $10\%-20\%$ smaller {\it effective} $v_{rot}$ value can easily
be envisaged.
There are also significant astrophysical uncertainties in this quantity.

\vskip 0.5cm
\centerline{\epsfig{file=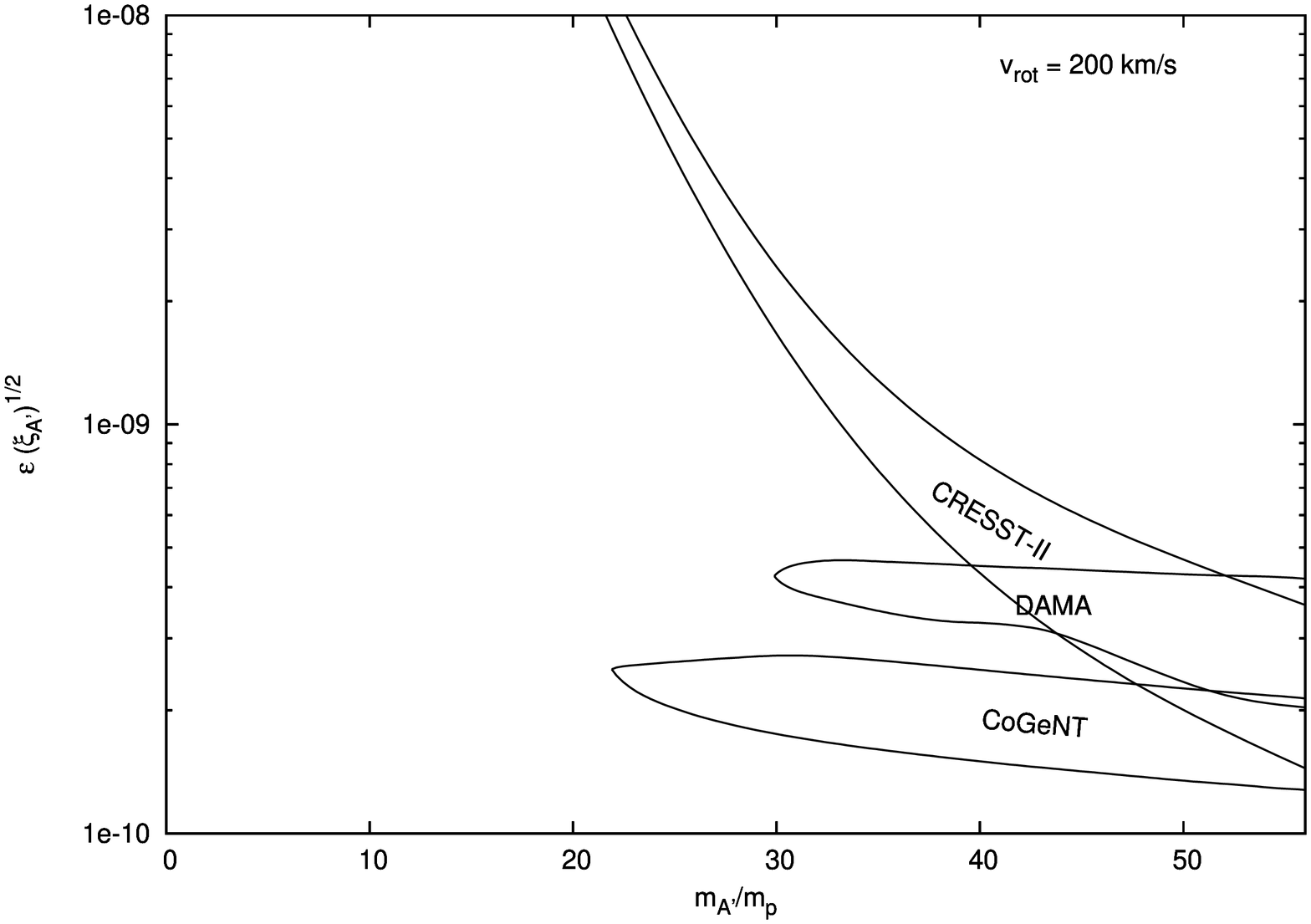,angle=270,width=13.0cm}}
\vskip 0.3cm
\noindent
{\small
Figure 2: DAMA, CoGeNT and CRESST-II favored regions of 
parameter space in the mirror dark matter model for
$v_{rot} = 200$ km/s.
}
\vskip 0.6cm

The 90\% (99\%) C.L. favored region of parameter 
space is bounded by the contours where 
$\chi^2 (m_{A'}, \epsilon \sqrt{\xi_{A'}}) = \chi^2_{min} + 4.6 (9.2)$. 
In figure 2 we plot the CRESST-II 90\% C.L. favored region along
with the 99\% C.L. favored region of parameter space for DAMA and
CoGeNT\footnote{
The CoGeNT experiment does not discriminate electron recoils from
nuclear recoils.
In principle, e$'$ inelastic scattering on bound electrons in the target
can therefore also
contribute to the signal at low energies $E_R \stackrel{<}{\sim}$ keVee.
Previous work\cite{footelec,foot2} has indicated that the e$'$
contribution can be comparable
to the rate of nuclear recoils. However, the previous
analysis\cite{footelec,foot2}
assumed that the flux of e$'$ at the detector was much larger than the
flux of mirror nuclei
due to the large dispersion velocity $v_0 [e'] \gg v_{rot}$. In reality
this cannot be the case.
A larger flux of e$'$ would lead to a greater rate of e$'$ capture in
the Earth c.f. mirror nuclei capture,
and hence this would lead to an increasing mirror electric charge within
the Earth, $Q'_E$.
In fact, we expect $Q'_E$ to be generated such that the flux of e$'$ and 
mirror nuclei at the Earth's surface 
approximately equalize.
Taking this effect into account reduces the expected e$'$ flux by more 
than an order of magnitude,
leaving nuclear recoils as the dominant contribution to the rate.}. 


Figure 2 indicates that in the high
mass region, $m_{A'} \stackrel{>}{\sim} 50 m_p$ including $A' \sim$
Fe$'$, DAMA
encompasses lower $\epsilon \sqrt{\xi_{A'}}$ values. 
This is because dark matter scattering on Iodine  becomes
kinematically favored and can dominate over Sodium scattering in this
parameter region. The resulting
lower values of $\epsilon \sqrt{\xi_{A'}}$ produce a favored region of
parameter space 
overlapping with that of the other experiments.
If the spectrum of mirror dark matter elements has some resemblance to
ordinary matter,
we might expect Fe$'$ to be the dominant heavy mirror metal component.
To study this
possibility further
we fix $A' =$ Fe$'$ ($m_{A'} = 55.8m_p$) and vary $v_{rot}$ in the
range: $160 \le v_{rot} ({\rm km/s}) \le 250$.  
The resulting 
$\chi^2_{min}$ values are given in table 3. Expanding around these
$\chi^2_{min}$ values we 
obtain
the CRESST-II 90\% C.L. and the DAMA, CoGeNT 99\% C.L. favored parameter 
regions in figure 3.

\begin{table}
\centering
\begin{tabular}{c c c}
\hline\hline
Experiment &  $\chi^2 (min)/d.o.f. $ & Best fit parameters  \\
\hline
DAMA (annual mod.) &   $5.5/10$ &  $v_{rot} = 210$ km/s  , 
\ $\epsilon \sqrt{\xi_{A'}}/10^{-10} = 3.1 $  \\
CoGeNT (spectrum) &   $15.8/12$  & $v_{rot} = 201$ km/s  , 
\ $\epsilon \sqrt{\xi_{A'}}/10^{-10} = 1.6 $  \\
CRESST (spectrum) &   $0.3/3$ &  $v_{rot} = 250$ km/s  , 
\ $\epsilon \sqrt{\xi_{A'}}/10^{-10} = 1.7 $  \\
\hline\hline
\end{tabular}
\caption{Summary of $\chi^2 (min)$ for the relevant data sets from the 
DAMA, CoGeNT and CRESST-II experiments for $A' =$ Fe$'$.}
\end{table}

\vskip 0.5cm
\centerline{\epsfig{file=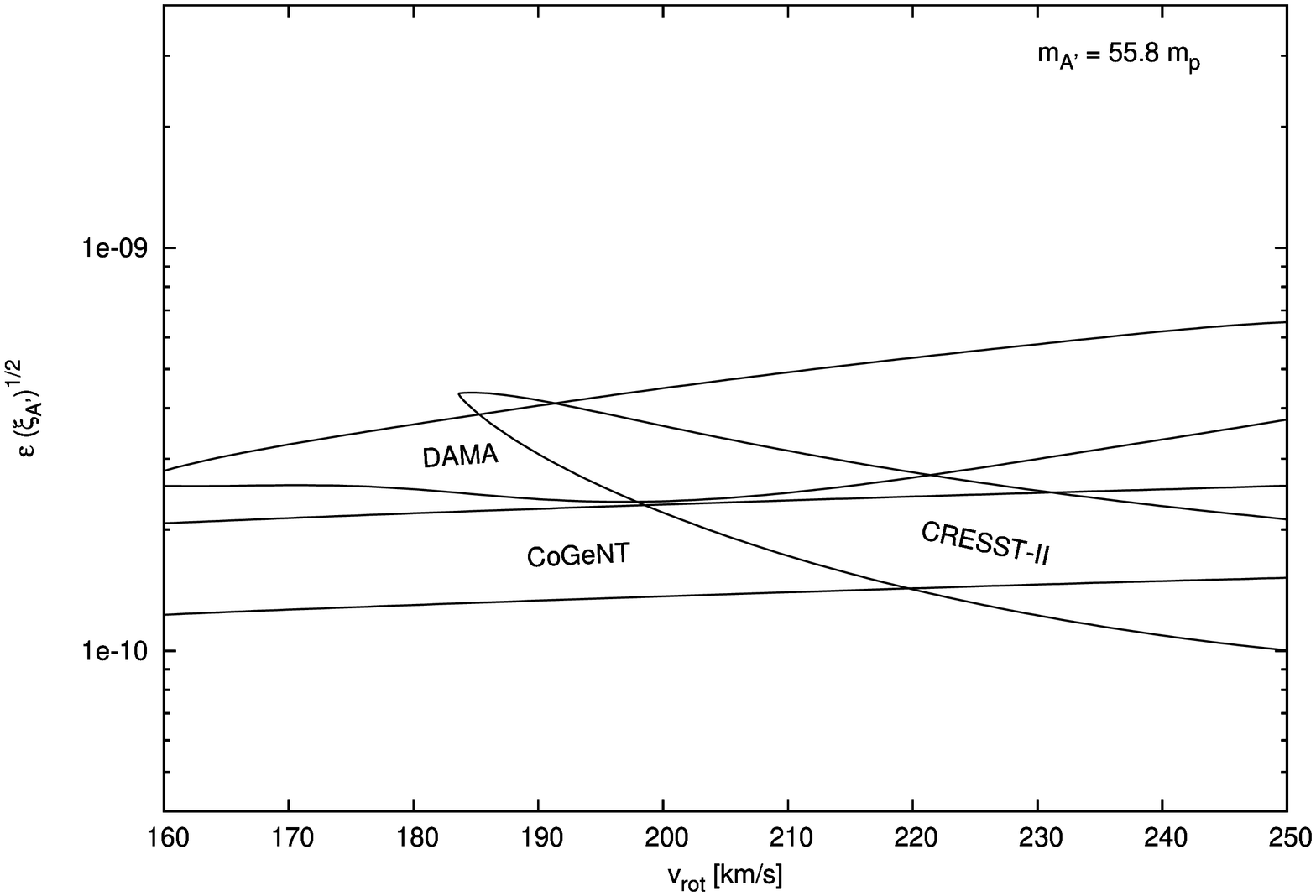,angle=270,width=14cm}}
\vskip 0.4cm
\noindent
{\small
Figure 3: DAMA, CoGeNT and CRESST-II favored regions of 
parameter space in the mirror dark matter model for 
$A' = $ Fe$'$ (i.e. $m_{A'} = 55.8m_p$).
}

\vskip 0.8cm

\section{Two examples, P1 and P2} 

As discussed earlier, the systematic uncertainties are potentially quite
significant for CoGeNT.
The favored region can move up or down by  $\sim 20\%$ as future data is
collected and backgrounds
are better understood.
A small channeling fraction or quenching factors outside the range
considered in Eq.(\ref{moomba}),
can also move the DAMA region somewhat.  In this light,
figures 2,3 offer some encouragement as they indicate some parameter
space where all three 
experiments can be explained by Fe$'$ scattering.
At this point one could combine all three of these experiments and do a
fit 
using the combined $\chi^2$.
However, in view of the potentially
large systematic uncertainties it is not clear how useful
such a combined analysis would be.  Instead we consider two
example reference points. The first one is located near the overlapping 
allowed regions indicated in figure 2:
\begin{eqnarray}
P1: A' &=& {\rm Fe}' \ (m_{Fe'} \simeq 55.8m_p),\ v_{rot} = 200\ {\rm
km/s}, \ 
\epsilon \sqrt{\xi_{Fe'}} = 2.2\times 10^{-10}\ .
\label{p1}
\end{eqnarray}
Figure 4a,b,c, compares the predicted rates assuming the $P1$ parameter
set
with the relevant  DAMA, CoGeNT and CRESST-II data.

\vskip 0.5cm
\centerline{\epsfig{file=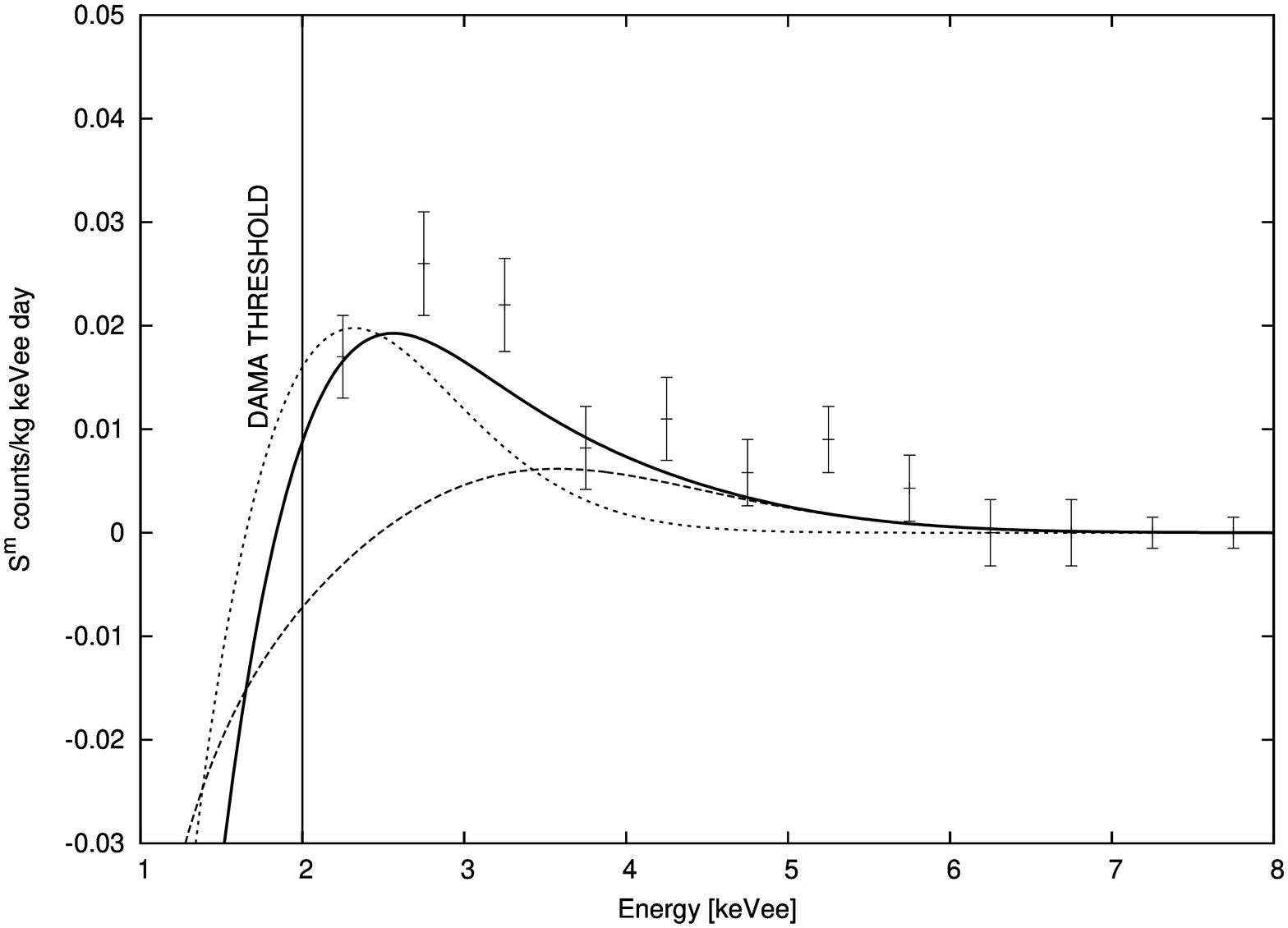,angle=270,width=14cm}}
\vskip 0.3cm
\noindent
{\small
Figure 4a:
DAMA annual modulation spectrum for mirror dark matter 
with parameter choice $P1$ (solid line). 
The separate contributions from dark matter scattering off Sodium
(dashed line) and Iodine (dotted line) 
are shown.   In this example $q_{Na} = 0.33, \ q_I = 0.20$.} 
\vskip 0.5cm
\centerline{\epsfig{file=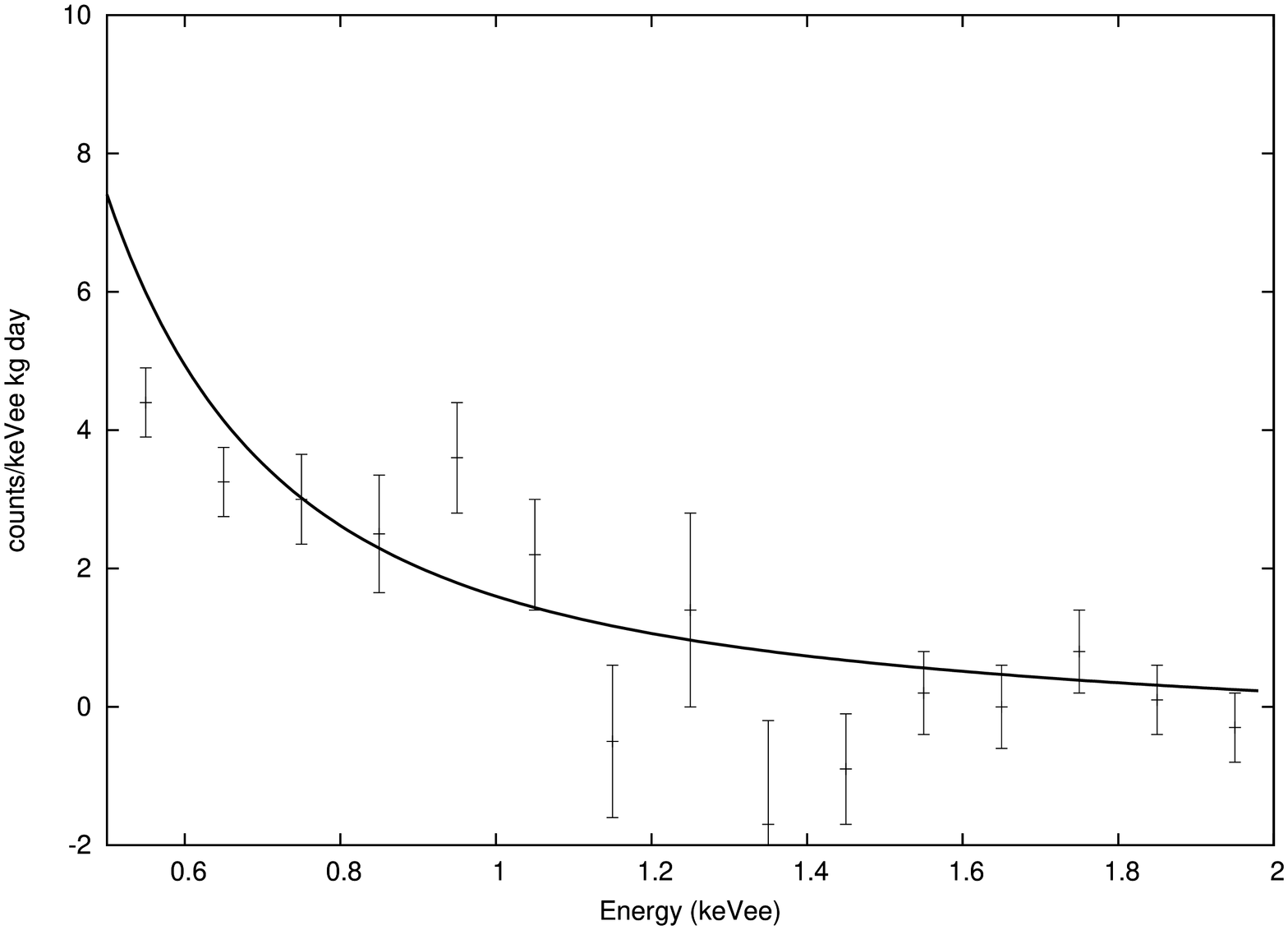,angle=270,width=13.7cm}}
\vskip 0.3cm
\noindent
{\small
Figure 4b: CoGeNT spectrum for mirror dark matter with 
parameters $P1$ (solid line).
In this example $q_{Ge} = 0.17$.
}

\vskip 0.5cm
\centerline{\epsfig{file=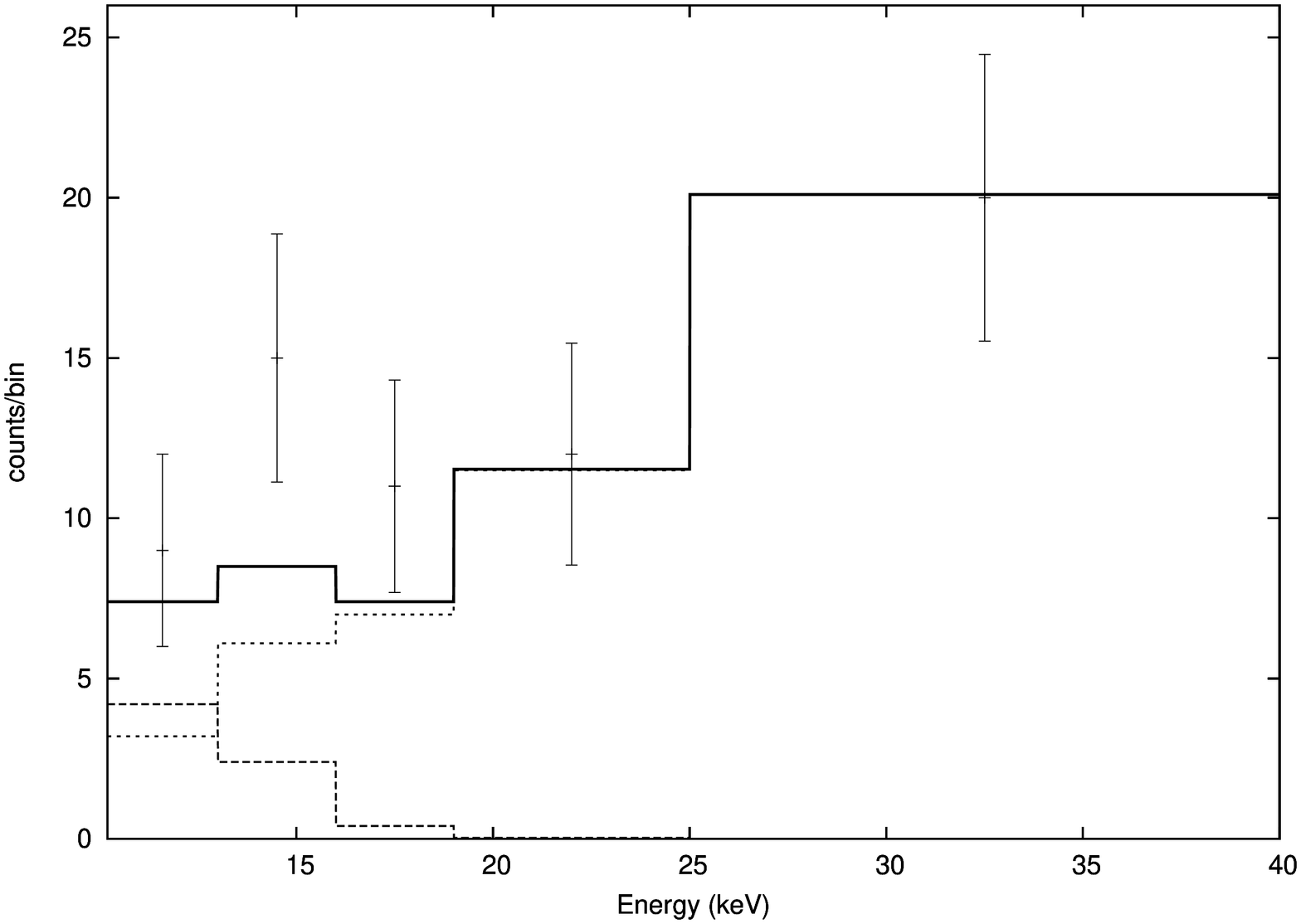,angle=270,width=13.7cm}}
\vskip 0.3cm
\noindent
{\small
Figure 4c: CRESST-II spectrum for mirror dark matter with 
parameters $P1$ (solid line). The signal component (dashed line) and
background component (dotted line) are also shown.
}
\vskip 0.8cm

For the example point $P1$, the CRESST-II signal is dominated by Fe$'$
scattering on the
Ca target element. Of course the reasons for this are purely kinematical
being due to
$m_{Fe'} \sim m_{Ca}$.

The change in sign of the annual modulation amplitude at 
low energies indicated in figure 4a 
will not necessarily hold if there are 
lighter more abundant components such as O$'$, Ne$'$. 
The positive contributions to the annual modulation amplitude from these
lighter 
components can overwhelm the
negative contribution due to the heavier components at low energies.
We illustrate this point with the following example:
\begin{eqnarray}
P2: A' = {\rm Fe}', \ \epsilon\sqrt{\xi_{Fe'}} = 1.7 \times 10^{-10},
\ \xi_{O'}/\xi_{Fe'} = 10,
\ v_{rot} = 210 \ {\rm km/s}
\ .
\end{eqnarray}
Figure 5a,b,c, compares the predicted rates for the reference point $P2$
with the relevant  DAMA, CoGeNT and CRESST-II data. 

\vskip 0.5cm
\centerline{\epsfig{file=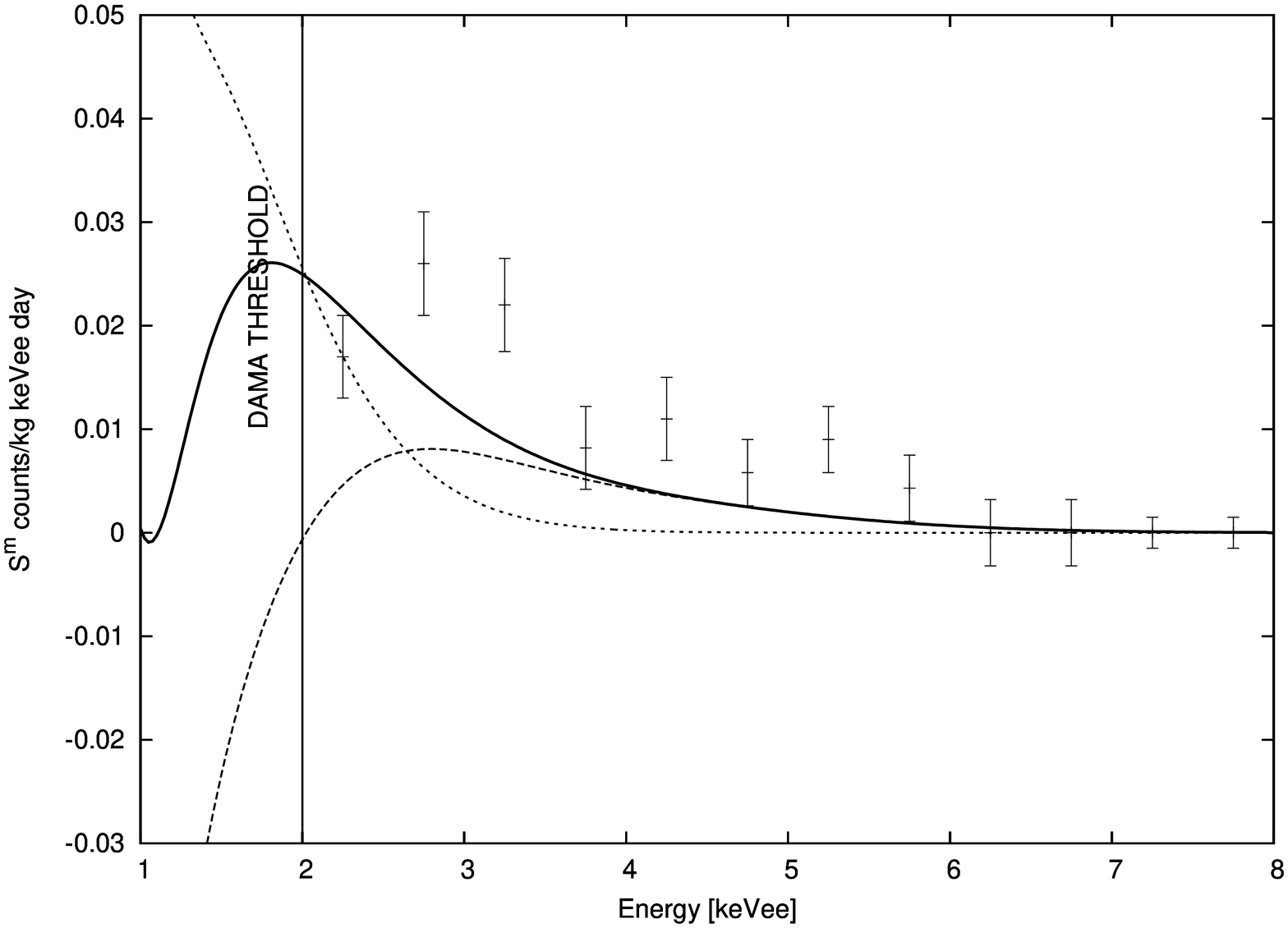,angle=270,width=14.0cm}}
\vskip 0.3cm
\noindent
{\small
Figure 5a:
DAMA annual modulation spectrum for mirror dark matter with 
parameters $P2$ (solid line). The separate contributions from
Fe$'$ interactions (dashed line) and O$'$ (dotted line) interactions are
also shown.
In this example $q_{Na} = 0.34$, $q_I = 0.20$.
}

\vskip 0.5cm
\centerline{\epsfig{file=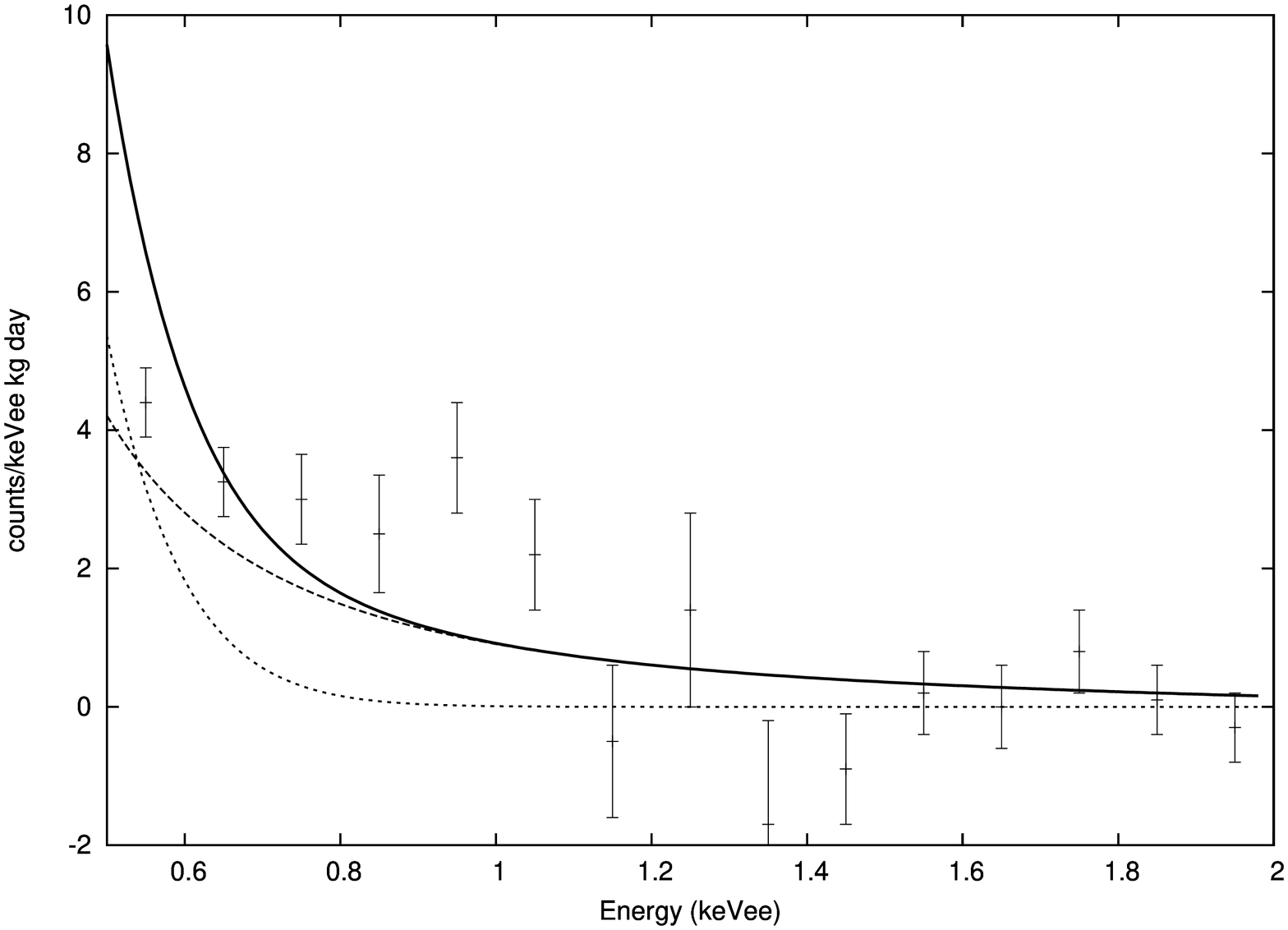,angle=270,width=14.0cm}}
\vskip 0.3cm
\noindent
{\small
Figure 5b: 
CoGeNT spectrum for mirror dark matter with 
parameters $P2$ (solid line). The separate contributions from
Fe$'$ interactions (dashed line) and O$'$ (dotted line) interactions are
also shown. 
In this example $q_{Ge} = 0.17$.}

\vskip 0.2cm
\centerline{\epsfig{file=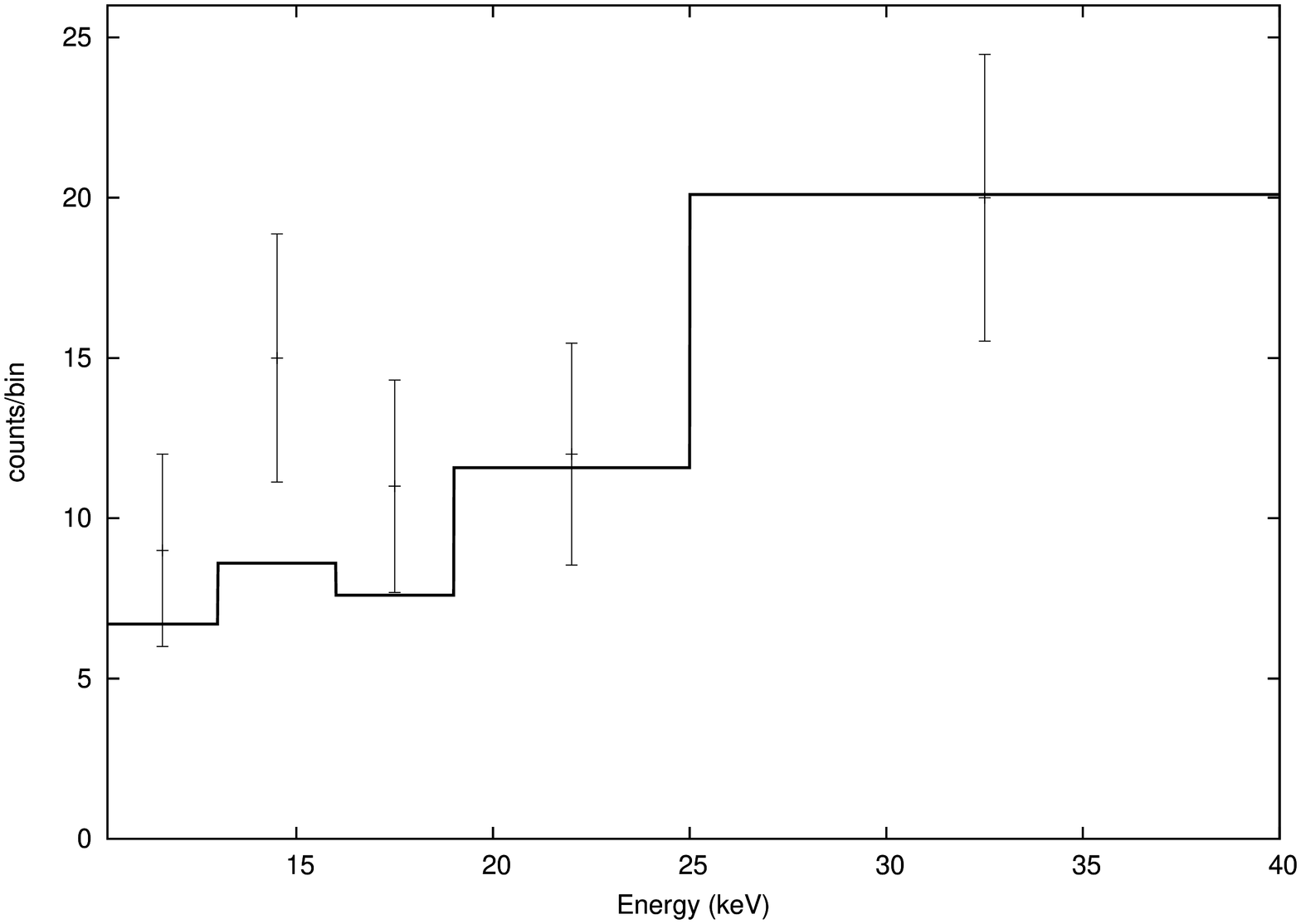,angle=270,width=14.0cm}}
\vskip 0.3cm
\noindent
{\small
Figure 5c: CRESST-II spectrum for mirror dark matter with 
parameters $P2$ (solid line).}

\vskip 0.6cm

Figures 5 show that with low $v_{rot} \stackrel{<}{\sim} 220$ km/s  
light metal components ($\sim $ O$'$) and lighter components (H$'$,
He$'$) can be
kinematically suppressed in DAMA, CoGeNT and CRESST-II. 
Future experiments 
at lower thresholds, such as Texono\cite{texono2},
should be able
to probe such light components. 

Observe that the dark matter signal contribution in CRESST-II is
relatively 
low in these examples. However it could easily be enhanced in many ways. 
Increasing $\epsilon \sqrt{\xi_{A'}}$  will increase the rate in
CRESST-II and improve the fit in DAMA and
is therefore an obvious possibility. 
Another possibility is that 
the CRESST-II energy scale might be overestimated by e.g. $\sim 10\%$. 
If CRESST-II events are at lower energies
then this can increase the expected rate in CRESST-II.
Also, a slight increase in $\bar m$ of order $\sim 20\%$ can greatly
improve the fit in CRESST-II.
Such small changes of $\bar m$ do not significantly affect DAMA or
CoGeNT since these
experiments are sensitive to dark matter in the body of the Maxwellian
distribution\cite{foot2}
while CRESST-II is probing interactions in the tail of the dark matter
velocity distribution (in this example).


\vskip 0.5cm
\centerline{\epsfig{file=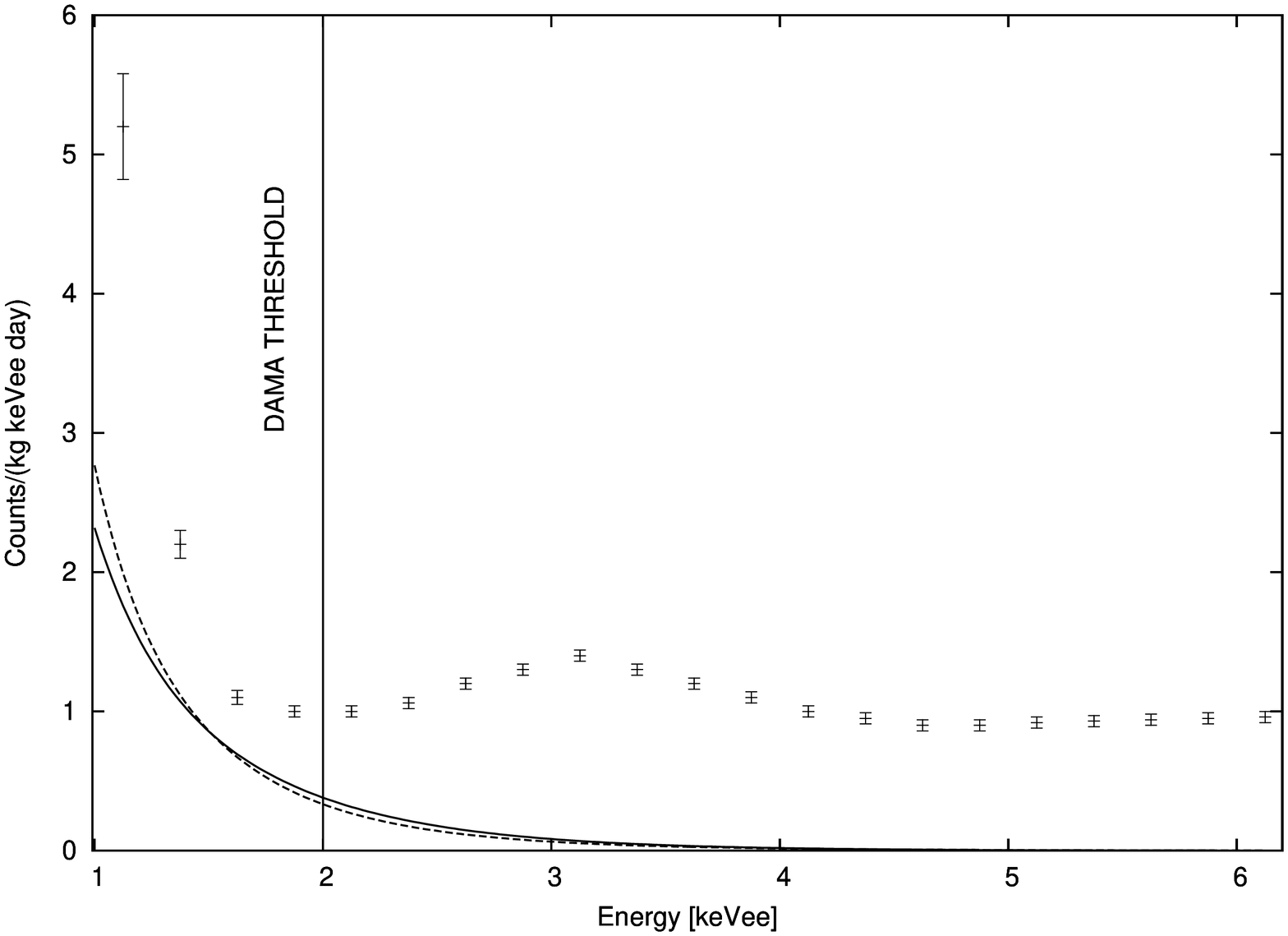,angle=270,width=14cm}}
\vskip 0.3cm
\noindent
{\small
Figure 6:
DAMA spectrum for mirror dark matter with 
parameters $P1$ (solid line), $P2$ (dashed line). 
In this example $q_{Na} = 0.34$, $q_I = 0.20$.}
\vskip 0.7cm

Future data from DAMA, CoGeNT, CRESST-II and other experiments will
obviously be able
to constrain the parameter space within the mirror dark matter
framework.
As discussed recently\cite{diurnal},
a particularly striking diurnal modulation signal should be observable
for a detector located in the southern hemisphere, and maybe even in 
a detector in the northern hemisphere at low
latitudes, such as detectors in Jin-Ping Underground laboratory.
In the meantime, we must rely on annual modulation and spectrum data.
In figure 6 we give the predicted spectrum for DAMA/Libra and Figure 7
the predicted
annual modulation spectrum in CoGeNT for each reference point, 
$P1,\ P2$ [detection efficiency = 1 for
these figures]. Clearly, the initial annual modulation amplitude
measured by CoGeNT
to be $A \approx 0.46 \pm 0.17$ cpd/kg/keVee averaged over the energy
range: 
$0.5 < E({\rm keVee}) < 3.0$, is much larger than 
that predicted by our example points.
It will be interesting to see if CoGeNT's hint of a large annual
modulation amplitude is confirmed by
CDEX, C-4, and other experiments.

\vskip 0.5cm
\centerline{\epsfig{file=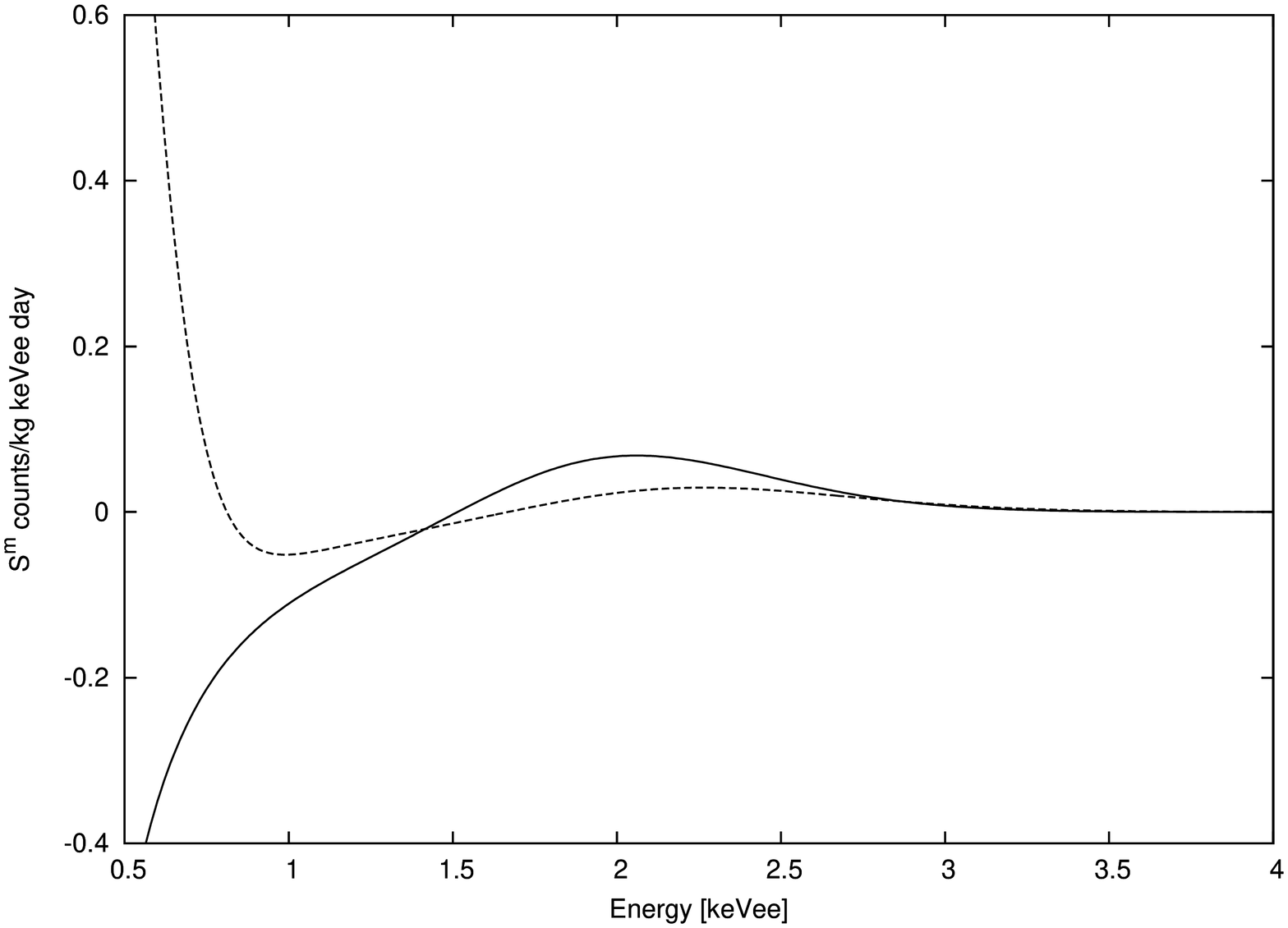,angle=270,width=14cm}}
\vskip 0.3cm
\noindent
{\small
Figure 7:
CoGeNT annual modulation spectrum for mirror dark matter with 
parameters $P1$ (solid line), $P2$ (dashed line). 
In this example $q_{Ge} = 0.17$.
}
\vskip 0.8cm

Let us make a few comments.
Firstly, there are alternative explanations of DAMA, CoGeNT and 
CRESST-II within the mirror dark matter framework.
For example, if $v_{rot}$ is high ($\stackrel{>}{\sim}$ 240 km/s) then
DAMA and CoGeNT can be mainly 
detecting the lighter
metal components $\sim $ O$'$-Si$'$ \cite{foot2,foot1}. In this case, it
is possible that
CRESST-II might be seeing a less abundant heavier component (e.g.
Fe$'$), or even the same component if 
$v_{rot}$ is very high $\stackrel{>}{\sim} 260$ km/s.
Secondly, more generic hidden sector dark matter can also explain the
data
in a similar way to the mirror dark matter case. 
One simply requires a hidden sector with two or more stable particles
charged under
an unbroken $U(1)'$ gauge interaction which is
kinetically mixed with standard $U(1)_Y$\cite{foot1,foot2008}.
These ingredients
can arise, for example, in spontaneously broken mirror
models\cite{chinese}.  

Finally, observe that
the experiments constrain $\epsilon \sqrt{\xi_{A'}}$ but do not allow
for a precise
determination of $\epsilon$ or $\xi_{A'}$ separately. 
It is difficult to estimate $\xi_{A'}$. Mirror metals might be generated 
in mirror stars, presumably at an early epoch\cite{mirrorstarstudy}. It
might also be possible
to produce the metal component in the early Universe, although this is
disfavored in
the simplest scenarios with high reheating temperature\cite{paolo2}.
A third possibility is that the average metal component is actually very
low
in the galaxy, $\langle \xi_{A'} \rangle \ll 10^{-2}$, but
its abundance is greatly enhanced in the galactic disk. One possible
mechanism for this
is the influence of the galactic magnetic field\cite{kolb} which can
potentially
exclude (or suppress) light elements ($m_{A'} \stackrel{<}{\sim}
m_{O'}$)
from entering the disk\cite{footz} (see also ref.\cite{zurek}).
This could thereby enhance the proportion of heavier components whose
abundance
should increase until the pressure equalizes with that of the rest of
the halo (at the same radial distance).
In this case $\xi_{A'} \sim 1$ is possible within the disk, i.e. at
the detector's location.
The above possibilities suggest a broad range for $\xi_{A'}$: $10^{-4}
\stackrel{<}{\sim} \xi_{A'} \stackrel{<}{\sim}
1$. If the direct detection experiments are due to mirror dark matter
with $\epsilon \sqrt{\xi_{A'}} \sim
\ {\rm few}\ 10^{-10}$
then this suggests that $\epsilon$ is in the range\footnote{
Similar values of $\epsilon$ can be motivated from an astrophysical
argument. Mirror dark matter is dissipative and radiative cooling would
cause the halo to collapse 
on a time scale of a few hundred million years unless a substantial heat
source
exists. Kinetic mixing induced
processes in the core of ordinary supernova 
(such as $e' \bar e'$ production via plasmon decay\cite{raf}) can
supply the energy needed to
compensate for the energy lost from the halo due to radiative cooling if
$\epsilon \sim 10^{-9}$\cite{sph,fsfsfs}.}
$\sim 10^{-10} - 10^{-8}$. 
The most stringent laboratory limit on $\epsilon$ arises 
from invisible decays of orthopositronium\cite{gla,ortho}, $\epsilon
\stackrel{<}{\sim} 1.5 \times 10^{-7}$.
An important proposal exists\cite{ortho2} for a more sensitive
orthopositronium experiment which
can cover much of the $\epsilon$ range of interest. Also, note that
values of $\epsilon \stackrel{>}{\sim} 10^{-9}$ can also be probed in
early 
Universe cosmology\cite{footnov}. 
In particular, forthcoming results expected from the Planck mission
might
shed some light on $\epsilon$ in the near future. There are also other
interesting possible effects
of the kinetic mixing\cite{review,mitra}.

\section{Constraints from CDMS and XENON100}

Let us now examine the question of the compatibility of this model with
the
constraints from the XENON100\cite{xenon100}, 
CDMS/Ge\cite{cdmsge} and CDMS/Si\cite{cdmssi}
experiments\footnote{
There are also lower threshold analysis by the XENON10\cite{xenon10} and 
CDMS collaborations\cite{cdms}.
However it has been argued\cite{collarguts} that neither analysis can
exclude light dark matter when systematic uncertainties are properly
taken into account.}.
This question is somewhat non-trivial since the answer depends
sensitively on the
recoil energy threshold which (typically) has at least $20\%$
uncertainty and often
the subject of controversy. The XENON100 experiment,
which has the largest exposure,
has especially poor knowledge of energy calibration in the important low
energy region
(see ref.\cite{collarzzz,damaguts} for relevant discussions).
In view of the above, we explore the compatibility question by
estimating the energy threshold
for which the parameter point $P1$ can be excluded at $95\%$ C.L. [The
constraint on $P2$ 
is similar to that of $P1$].  
We have taken into account the relevant
detection efficiencies, exposure time etc, although we conservatively
ignore effects
of detector resolution for the XENON100 experiment.
Our results are given in table 4. 
Note that the $95\%$ C.L. limits in the second column arise since
CDMS/Ge, CDMS/Si and XENON100 observed $2, 0, 1$ events respectively
in the low energy region. 

\begin{table}
\centering
\begin{tabular}{c c c c c}
\hline\hline
Experiment & $95\%$ C.L limit & $E_R^{est}$ threshold & $E_R^{nom}$
threshold & $100\left[ 1 - {E_R^{nom} \over E_R^{est}}\right]$ \\
\hline
CDMS/Ge & 6.3  & 13.1 keV & 10.0 keV & 23.7\%    \\

CDMS/Si & 3.0 & 8.5  keV & 7.0 keV & 17.6 \% \\

XENON100 & 4.7 & 12.9 keV & 8.4 keV  & 34.9 \%    \\
\hline\hline
\end{tabular}
\caption{Estimated energy threshold ($E_R^{est}$) for which the 
CDMS/Ge, CDMS/Si and XENON100 experiments 
are consistent with the mirror dark matter expectations assuming the
example point, $P1$.
$E_R^{nom}$ is the nominal energy threshold of each experiment. The last
column gives
the percentage at which the energy threshold needs to have been
underestimated for
the point $P1$ to be consistent with these experiments at $95\%$ C.L.}
\end{table}

Table 4 indicates that mirror dark matter with parameters given by the
example $P1$ is
consistent with the CDMS/Ge, CDMS/Si and XENON100 data provided that the
energy
scale has been underestimated by of order $20-30\%$ respectively.
This level of energy scale uncertainty is within critical assessments of
these
experiments discussed in the literature. 
We conclude that the considered mirror dark matter explanation of the
DAMA, 
CoGeNT and CRESST-II data is  consistent
with the data from CDMS/Si, CDMS/Ge and XENON100 experiments when
reasonable systematic
uncertainties in energy scale are included.
It should also be noted that the two events seen in CDMS/Ge
and also the events seen just above the threshold in the Edelweiss
experiment\cite{edelweiss} 
are both compatible with dark matter interactions within this model
given these energy scale uncertainties.

Finally, the KIMS experiment\cite{kims} is currently running with a 
$\sim 100$ kg CsI target. They have reported
a (2$\sigma$) limit on the rate of dark matter interactions from a Pulse
Shape Analysis (PSA) of 
their spectrum: $R < 0.02$ cpd/kg/keVee  for $ 3.0 < E_R ({\rm keVee})
< 4.0$.
We find that the Iodine interaction rate in a CsI target 
predicted by mirror dark matter for
the particular point $P1$ is
$R \approx 0.006$ cpd/kg/keVee in the $3 < E_R({\rm keVee}) < 4$
bin and 
falling sharply at higher energies. [Assuming
here that KIMS has the same resolution and $q_I$ quenching factor as the
DAMA/Libra experiment]. Thus,
their PSA does not constrain mirror dark matter.  However, the KIMS
experiment 
also aims to measure/constrain the annual
modulation amplitude.
The annual modulation results are anticipated to be reported within the
coming year.
Assuming that they can reach a similar energy threshold to DAMA, they
can potentially
see a positive signal for a signficant portion of parameter space (c.f dotted line in figure 4a).

\section{Conclusion}

In conclusion,
we have examined the new results from the CRESST-II experiment along
with the latest
results from DAMA and CoGeNT experiments in the context of the mirror
dark matter framework.
In this framework
dark matter consists of a spectrum of mirror elements: H$'$, He$'$,
O$'$, Fe$'$, ... of known
masses.
Under the simplifying but reasonable assumption that DAMA, CoGeNT and
CRESST-II might
be observing a particular dark matter component, $A'$,
we have found that mirror dark matter models can explain the data from
each experiment.
In particular, we have found that each experiment can be explained by
$A' \sim $ Fe$'$ interactions
if $\epsilon \sqrt{\xi_{Fe'}} \approx 2 \times 10^{-10}$ and 
$v_{rot} \sim 200$ km/s.
Other regions of parameter space are possible.
We have also shown that the considered
explanation is consistent with the results of the other experiments when 
reasonable systematic uncertainties in energy scale are considered.

\vskip 2.9cm
\noindent
{\large \bf Acknowledgments}

\vskip 0.2cm
\noindent
The author would like to thank 
Juan Collar, Juhee Lee and Leo Stodolsky 
for helpful correspondence.
This work was supported by the Australian Research Council.


\begin{thebibliography}{999}

\bibitem{dama1}
R. Bernabei {\it et al}. (DAMA Collaboration), 
Riv. Nuovo Cimento. {\bf 26}, 1 (2003) [astro-ph/0307403]; Int. J. Mod.
Phys. E{\bf 13}, 2127 (2004); Phys. Lett. B{\bf 480}, 23 (2000).

\bibitem{dama2}
R. Bernabei {\it et al}. (DAMA Collaboration), 
Eur. Phys. J. C{\bf 67}, 39 (2010) 
[arXiv: 1002.1028];
Eur. Phys. J. C{\bf 56}, 333 (2008) [arXiv:0804.2741]. 



\bibitem{dm}
A. K. Drukier, K. Freese and D. N. Spergel, Phys. Rev. D{\bf 33}, 3495
(1986);
K. Freese, J. A. Frieman and A. Gould, Phys. Rev. D{\bf 37}, 3388
(1988).
 

\bibitem{damamuon}
R.~Bernabei {\it et al.},
arXiv:1202.4179.




\bibitem{cogent}
C. E. Aalseth {\it et al.} (CoGeNT Collaboration),  Phys. Rev. Lett.
{\bf 106}, 131301 (2011)
[arXiv:1002.4703]; Phys. Rev. Lett. {\bf 107}, 141301 (2011) [arXiv:
1106.0650].

\bibitem{cresst-II}
G.~Angloher, M.~Bauer, I.~Bavykina, A.~Bento, C.~Bucci, C.~Ciemniak,
G.~Deuter and F.~von Feilitzsch {\it et al.},
arXiv:1109.0702.

\bibitem{std}
 C.~Savage, G.~Gelmini, P.~Gondolo and K.~Freese,
JCAP {\bf 0904}, 010 (2009)
[arXiv:0808.3607];
C.~Savage, G.~Gelmini, P.~Gondolo and K.~Freese,
Phys.\ Rev.\ D {\bf 83}, 055002 (2011)
[arXiv:1006.0972];
Y.~Mambrini,
JCAP {\bf 1009}, 022 (2010)
[arXiv:1006.3318];
 D.~Hooper, J.~I.~Collar, J.~Hall, D.~McKinsey and C.~Kelso,
Phys.\ Rev.\ D {\bf 82}, 123509 (2010)
[arXiv:1007.1005];
P.~J.~Fox, J.~Liu and N.~Weiner,
Phys.\ Rev.\ D {\bf 83} (2011) 103514
[arXiv:1011.1915];
J.~L.~Feng, J.~Kumar, D.~Marfatia and D.~Sanford,
Phys.\ Lett.\ B {\bf 703}, 124 (2011)
[arXiv:1102.4331];
C.~Arina, J.~Hamann and Y.~Y.~Y.~Wong,
JCAP {\bf 1109}, 022 (2011)
[arXiv:1105.5121]; 
D.~Hooper and C.~Kelso,
Phys.\ Rev.\ D {\bf 84}, 083001 (2011)
[arXiv:1106.1066];
P.~Belli, R.~Bernabei, A.~Bottino, F.~Cappella, R.~Cerulli, N.~Fornengo
and S.~Scopel,
Phys.\ Rev.\  D {\bf 84}, 055014 (2011) [arXiv:1106.4667];
T.~Schwetz and J.~Zupan,
JCAP {\bf 1108}, 008 (2011) [arXiv:1106.624];
M.~Farina, D.~Pappadopulo, A.~Strumia and T.~Volansky,
JCAP {\bf 1111}, 010 (2011)
[arXiv:1107.0715];
P.~J.~Fox, J.~Kopp, M.~Lisanti and N.~Weiner,
Phys.\ Rev.\ D {\bf 85}, 036008 (2012)
[arXiv:1107.0717];
C.~McCabe,
Phys.\ Rev.\  D {\bf 84}, 043525 (2011) [arXiv:1107.0741];
N.~Fornengo, P.~Panci and M.~Regis,
Phys.\ Rev.\  D {\bf 84}, 115002 (2011)
[arXiv:1108.4661];
J.~M.~Cline and A.~R.~Frey,
Phys.\ Lett.\ B {\bf 706}, 384 (2012)
[arXiv:1109.4639];
J.~Kopp, T.~Schwetz and J.~Zupan,
arXiv:1110.2721];
M. ~T. ~Frandsen, F.~Kahlhoefer, C.~McCabe, S.~Sarkar and
K.~Schmidt-Hoberg,
JCAP {\bf 1201}, 024 (2012) [arXiv:1111.0292];
P.~Gondolo and G.~B.~Gelmini,
arXiv:1202.6359.


\bibitem{foot69}
R. Foot, Phys. Rev. D{\bf 69}, 036001 (2004) [hep-ph/0308254]. 





\bibitem{footold}
R. Foot,  Mod. Phys. Lett. A{\bf 19},
1841 (2004) [astro-ph/0405362]; astro-ph/0403043; Phys. Rev. D{\bf 74},
023514 (2006) [astro-ph/0510705].

\bibitem{foot2008}
R. Foot, Phys. Rev. D{\bf 78}, 043529 (2008) [arXiv: 0804.4518].


\bibitem{foot2}
R. Foot, Phys. Rev. D{\bf 82}, 095001 (2010) [arXiv: 1008.0685];
Phys. Lett. B{\bf 692}, 65 (2010) [arXiv: 1004.1424].

\bibitem{foot1}
R. Foot, Phys.\ Lett.\ B {\bf 703}, 7 (2011) [arXiv:1106.2688].


\bibitem{collartaupxxx}
Talk by J. Collar, 
TAUP 2011 workshop, Munich, Germany Sep 5-9, 2011.


\bibitem{flv}
R. Foot, H. Lew and R. R. Volkas, Phys. Lett. B{\bf 272}, 67 (1991);
Mod. Phys. Lett. A{\bf 7}, 2567 (1992).

\bibitem{review}
R. Foot, Int. J. Mod. Phys. D{\bf 13}, 2161 (2004)
[astro-ph/0407623]; 
Int. J. Mod. Phys. A{\bf 19} 3807 (2004) [astro-ph/0309330];
P. Ciarcelluti, Int. J. Mod. Phys. D{\bf 19}, 2151 (2010) [arXiv:
1102.5530].

\bibitem{some}
H. M. Hodges, Phys. Rev. D{\bf 47}, 456 (1993); 
Z. Berezhiani, D. Comelli and F. L. Villante,
Phys. Lett. B{\bf 503}, 362 (2001) [hep-ph/0008105];
L. Bento and Z. Berezhiani, Phys. Rev. Lett. {\bf 87}, 231304 (2001)
[hep-ph/0107281]; 
A. Yu. Ignatiev and R. R. Volkas, Phys. Rev. D{\bf 68}, 023518 (2003)
[hep-ph/0304260];
R. Foot and R. R. Volkas, Phys. Rev. D{\bf 68}, 021304 (2003)
[hep-ph/0304261]; Phys. Rev. D{\bf 69}, 123510 (2004) [hep-ph/0402267];
Z. Berezhiani, P. Ciarcelluti, D. Comelli and F. L. Villante,
Int. J. Mod. Phys. D{\bf 14}, 107 (2005) [astro-ph/0312605];
P. Ciarcelluti, Int. J. Mod. Phys. D{\bf 14}, 187 (2005)
[astro-ph/0409630];
Int. J. Mod. Phys. D{\bf 14}, 223 (2005) [astro-ph/0409633].
For pioneering work, see: S. I. Blinnikov and M. Yu. Khlopov, Sov. J.
Nucl. Phys.
{\bf 36}, 472 (1981); Sov. Astron. {\bf 27}, 371 (1983).

\bibitem{he}
R. Foot and X-G. He, Phys. Lett. B{\bf 267}, 509 (1991). 

\bibitem{holdom} 
B.~Holdom,
Phys.\ Lett.\ B {\bf 166}, 196 (1986).

\bibitem{helm}
R. H, Helm, Phys. Rev. {\bf 104}, 1466 (1956). 

\bibitem{smith}
J. D. Lewin and P. F. Smith, Astropart. Phys. {\bf 6}, 87 (1996).


\bibitem{macho1}
Z.~K.~Silagadze,
Phys.\ Atom.\ Nucl.\  {\bf 60}, 272 (1997)
[Yad.\ Fiz.\  {\bf 60N2}, 336 (1997)] [hep-ph/9503481];
R.~Foot,
Phys.\ Lett.\ B {\bf 452}, 83 (1999) [astro-ph/9902065].

\bibitem{macho}
C.~Alcock {\it et al.}  [MACHO Collaboration],
Astrophys.\ J.\  {\bf 542}, 281 (2000) [astro-ph/0001272].
P.~Tisserand {\it et al.}  [EROS-2 Collaboration],
Astron.\ Astrophys.\  {\bf 469}, 387 (2007) [astro-ph/0607207].
  

\bibitem{sph}
R. Foot and R. R. Volkas,
Phys. Rev. D{\bf 70}, 123508 (2004) [astro-ph/0407522].


\bibitem{footz}
R. Foot, Phys. Lett. B{\bf 699}, 230 (2011) [arXiv:1011.5078].


\bibitem{bullet}
D.~Clowe, M.~Bradac, A.~H.~Gonzalez, M.~Markevitch, S.~W.~Randall,
C.~Jones and D.~Zaritsky,
Astrophys.\ J.\  {\bf 648}, L109 (2006)
[astro-ph/0608407].

\bibitem{zurab}
Z.~K.~Silagadze,
ICFAI U.\ J.\ Phys.\  {\bf 2}, 143 (2009)
[arXiv:0808.2595].

\bibitem{paolo2}
P. Ciarcelluti and R. Foot, Phys. Lett. B{\bf 690}, 462 (2010)
[arXiv:1003.0880].


\bibitem{newstudy}
N. Bozorgnia, G.B. Gelmini and P. Gondolo, JCAP {\bf 1011}, 019 (2010)
[arXiv: 1006.3110];
JCAP {\bf 1011}, 028 (2010) [arXiv: 1008.3676].



\bibitem{damares}
R.~Bernabei {\it et al.}  [DAMA Collaboration],
Nucl.\ Instrum.\ Meth.\  A {\bf 592}, 297 (2008) [arXiv:0804.2738].

\bibitem{tretyak}
V.~I.~Tretyak,
Astropart.\ Phys.\  {\bf 33}, 40 (2010) [arXiv:0911.3041].

\bibitem{kims}
S. K. Kim, on behalf of the KIMS collaboration,
TAUP 2011 workshop, Munich, Germany Sep 5-9, 2011.


\bibitem{hooper}
C.~Kelso, D.~Hooper and M.~R.~Buckley,
Phys.\ Rev.\  D {\bf 85}, 043515 (2012) [arXiv:1110.5338].

\bibitem{rot}
A.~Brunthaler, M.~J.~Reid, K.~M.~Menten, X.~-W.~Zheng, A.~Bartkiewicz,
Y.~K.~Choi, T.~Dame and K.~Hachisuka {\it et al.},
arXiv:1102.5350.

\bibitem{footelec}
R.~Foot,
Phys.\ Rev.\ D {\bf 80}, 091701 (2009)
[arXiv:0909.3126].


\bibitem{texono2}
Q.~Yue {\it et al.}  [for the CDEX-TEXONO Collaboration],
arXiv:1201.5373.

\bibitem{diurnal}
R.~Foot,
arXiv:1110.2908.


\bibitem{chinese}
J.~-W.~Cui, H.~-J.~He, L.~-C.~Lu and F.~-R.~Yin,
arXiv:1110.6893.

\bibitem{mirrorstarstudy}
Z.~Berezhiani, S.~Cassisi, P.~Ciarcelluti and A.~Pietrinferni,
Astropart.\ Phys.\  {\bf 24}, 495 (2006) [astro-ph/0507153].

\bibitem{kolb}
L.~Chuzhoy and E.~W.~Kolb,
JCAP {\bf 0907}, 014 (2009)
[arXiv:0809.0436].

\bibitem{zurek}
S.~D.~McDermott, H.~-B.~Yu and K.~M.~Zurek,
Phys.\ Rev.\ D {\bf 83}, 063509 (2011)
[arXiv:1011.2907 ].

\bibitem{raf}
G.~G.~Raffelt,
{\it  Chicago, USA: Univ. Pr. (1996) 664 p}.

\bibitem{fsfsfs}
R.~Foot and Z.~K.~Silagadze,
Int.\ J.\ Mod.\ Phys.\ D {\bf 14}, 143 (2005) [astro-ph/0404515].



\bibitem{gla}
S.~L.~Glashow,
Phys.\ Lett.\ B {\bf 167}, 35 (1986);
R.~Foot and S.~N.~Gninenko,
Phys.\ Lett.\ B {\bf 480}, 171 (2000) [hep-ph/0003278];
S.~V.~Demidov, D.~S.~Gorbunov and A.~A.~Tokareva,
Phys.\ Rev.\ D {\bf 85}, 015022 (2012)
[arXiv:1111.1072 ].

\bibitem{ortho}
A.~Badertscher, P.~Crivelli, W.~Fetscher, U.~Gendotti, S.~Gninenko,
V.~Postoev, A.~Rubbia and V.~Samoylenko {\it et al.},
Phys.\ Rev.\ D {\bf 75}, 032004 (2007)
[hep-ex/0609059].

\bibitem{ortho2}
P.~Crivelli, A.~Belov, U.~Gendotti, S.~Gninenko and A.~Rubbia,
JINST {\bf 5}, P08001 (2010) [arXiv:1005.4802].

\bibitem{footnov}
R.~Foot,
arXiv:1111.6366;
P.~Ciarcelluti and R.~Foot,
Phys.\ Lett.\ B {\bf 679}, 278 (2009)
[arXiv:0809.4438] and references there-in.


\bibitem{mitra}
R. Foot and S. Mitra, Astropart. Phys. {\bf 19}, 739 (2003)
[astro-ph/0211067]; 
Phys. Lett. A{\bf 315}, 178 (2003) [cond-mat/0306561];
Phys. Lett. B{\bf 558}, 9 (2003) [astro-ph/0301229];
S.~Mitra,
Phys.\ Rev.\ D {\bf 74}, 043532 (2006) [astro-ph/0605369].

\bibitem{xenon100}
E.~Aprile {\it et al.}  [XENON100 Collaboration],
Phys.\ Rev.\ Lett.\  {\bf 107},  131302 (2011)
[arXiv:1104.2549 ].

\bibitem{cdmsge}
Z.~Ahmed {\it et al.}  [The \ CDMS-II Collaboration],
\ Science {\bf 327}, \ 1619 \ (2010) [arXiv:0912.3592 ].


\bibitem{cdmssi}
J.~P.~Filippini, Ph. D. thesis (2008).

\bibitem{xenon10}
J.~Angle {\it et al.}  [XENON10 Collaboration],
Phys.\ Rev.\ Lett.\  {\bf 107},  051301 (2011)
[arXiv:1104.3088].

\bibitem{cdms}
Z.~Ahmed {\it et al.}  [CDMS-II Collaboration],
Phys.\ Rev.\ Lett.\  {\bf 106}, 131302 (2011)
[arXiv:1011.2482].
  
\bibitem{collarguts}
J.~I.~Collar,
arXiv:1103.3481 ;
arXiv:1106.0653 .


\bibitem{collarzzz}
J.~I.~Collar,
arXiv:1010.5187.


\bibitem{damaguts}
R.~Cerulli {\it et al.},
arXiv:1201.4582.

\bibitem{edelweiss}
E.~Armengaud {\it et al.}  [EDELWEISS Collaboration],
Phys.\ Lett.\  B {\bf 702}, 329 (2011)
[arXiv:1103.4070].




\end{thebibliography}
\end{document}